\documentstyle[preprint,prd,aps]{revtex}
\newcommand{\be}{\begin{equation}}
\newcommand{\ee}{\end{equation}}
\newcommand{\bea}{\begin{eqnarray}}
\newcommand{\eea}{\end{eqnarray}}
\begin{document}
\draft
\title{Nonrelativistic spin $\frac{1}{2}$ particle in an arbitrary non-Abelian magnetic field in two spatial dimensions}
\author{T.E. Clark\footnote{e-mail address: clark@physics.purdue.edu}, S.T. Love\footnote{e-mail address: love@physics.purdue.edu} and S.R. Nowling\footnote{e-mail address: nowling@uiuc.edu}}
\address{\it Department of Physics, 
Purdue University,
West Lafayette, IN 47907-1396}
\maketitle
\begin{abstract}
The (group and spin space) matrix Hamiltonian describing the dynamics of a nonrelativistic spin $1/2$ particle  moving in a static, but spatially dependent, non-Abelian magnetic field in two spatial dimensions is shown to take the form of an anticommutator of a nilpotent operator and its hermitian conjugate. Consequently, the (group space) matrix Hamiltonians for the two different spin projections form partners of a supersymmetric quantum mechanical system. The resulting supersymmetry algebra is exploited to explicitly construct the exact zero energy ground state wavefunction(s) for the system. The remaining eigenstates and eigenvalues of the two partner Hamiltonians form positive energy degenerate pairs. 
\end{abstract}

\newpage

The motion of a nonrelativistic spin 1/2 particle confined to move in a plane under the action of a magnetic field directed normal to the plane is a fundamental problem appearing in a variety of physical applications\cite{LL}-\cite{qhe}. Previously, we examined this problem for a static magnetic field having arbitrary spatial dependence on the planar coordinates. We showed\cite{CLN1}-\cite{CLN2} that the model exhibited a supersymmetry\cite{J}-\cite{R} which we consequently exploited to construct the exact zero energy normalizable ground state(s) for the system. 

In this note, we study an analogous problem involving the planar motion of the spin 1/2 particle under the action of a non-Abelian magnetic field which is also directed normal to the plane and again having arbitrary spatial dependence on the planar coordinates. Such a configration has also been argued\cite{Wen} to have relevance for various physical systems. The non-Abelian magnetic field strength is $\vec{B}^a=\hat{z}B^a(x,y)$, with $a$ denumerating the group generators, and 
\bea
B^a=F_{12}^a &=& \partial_1 A_2^a -\partial_2 A_1^a +\frac{g}{\hbar c} f_{abc} A_1^b A_2^c \\
&=&\epsilon_{ij}(\partial_i A_j^a +\frac{g}{2\hbar c} f_{abc} A_i^b A_j^c).
\eea
The indices $i,j~=~1,2$ label the spatial coordinates of the plane, while $g$ is the gauge charge and $f_{abc}$ are the group structure constants. Thus the fundamental representation matrices, $L^a$, satisfy $[L^a,L^b]=if_{abc}L^c$. It proves convenient to introduce matrix valued fields $B=L^a B^a$ and $A_i=L^aA_i^a$ so that
\be
B=\epsilon_{ij} (\partial_i A_j - \frac{ig}{\hbar c}A_i A_j) .
\ee 

In general, the vector potential can be decomposed into transverse and longitudnal pieces as 
\be
A_i=\epsilon_{ij}\partial_j K + \partial_i C~~;~~ C=L^aC^a~~,~~K=L^aK^a .
\ee
We choose to work in the Coulomb gauge defined by $\partial_i A_i =0$ and $C=0$ so that 
\be
A_i = \epsilon_{ij}\partial_j K .
\ee
Introducing the complex coordinates $x_{\pm}=x\pm iy$, the spatial components of the matrix valued vector potential and non-Abelian magnetic field strength take the form
\be
A_1=\partial_2 K = i(\partial_+ K-\partial_- K)
\ee
\be
A_2=-\partial_1 K = -(\partial_+ K+\partial_- K)
\ee
and
\be
B=-4\partial_+\partial_- K-\frac{2g}{\hbar c}[\partial_+ K  ,\partial_- K] .
\ee

The Hamiltonian governing the dynamics of a nonrelativistic spin $1/2$ fermion of mass $m$ carrying the fundamental representation of the gauge group moving in $2$ spatial dimensions under the influence of such a spatially dependent non-Abelian magnetic field is given by
\bea
H&=&\frac{1}{2m}(\frac{\hbar}{i}\partial_i - \frac{g}{c}A_i)^2 -\frac{g\hbar}{2mc}\sigma_3 B \cr
&=&\frac{1}{2m}\left[-4\hbar^2\partial_+\partial_- -\frac{4\hbar g}{c}(\partial_+ K\partial_- - \partial_- K \partial_+)+\frac{2g^2}{c^2}(\partial_+ K \partial_- K + \partial_- K \partial_+ K) \right.\cr
&&\left.\quad\quad +[\frac{4\hbar g}{c} \partial_+\partial_- K +\frac{2g^2}{c^2}(\partial_+ K \partial_- K - \partial_- K \partial_+ K)]\sigma_3 \right]\cr
&=&\left[\begin{array}{rr}H_{\uparrow}&0\\0&H_{\downarrow}\end{array}\right] ,
\eea
where $\sigma_3=\left[\begin{array}{rr}1&0\\0&-1\end{array}\right]$ is a Pauli matrix. 
Here  
\be
H_{\uparrow}=\frac{1}{2m}[2(\frac{\hbar}{i}\partial_+ - \frac{ig}{c}\partial_+ K)]
[2(\frac{\hbar}{i}\partial_- + \frac{ig}{c}\partial_- K)]=\frac{1}{2m}\pi_+\pi_- 
\ee
\be
H_{\downarrow}=\frac{1}{2m}[2(\frac{\hbar}{i}\partial_- + \frac{ig}{c}\partial_- K)][2(\frac{\hbar}{i}\partial_+ - \frac{ig}{c}\partial_+ K)]=\frac{1}{2m}\pi_-\pi_+
\ee
are the (group matrix) Hamiltonians for spin projections $+1/2 (-1/2)$ respectively, with
\be
\pi_{\mp}=2(\frac{\hbar}{i}\partial_{\mp} \pm \frac{ig}{c}\partial_\mp K)=\pi_{\pm}^\dagger 
\ee
satisfying $[\pi_+,\pi_-]=-\frac{2\hbar g}{c}B$.

This matrix Hamiltonian can further be written as the square of a hermitian operator $\cal Q$ as
\be
H = \frac{1}{2} {\cal Q}^2 ~;~  {\cal Q} = \frac{1}{\sqrt{m}}\pmatrix{ 0& -i \pi_+ \cr
i\pi_- & 0} = ({\cal Q})^\dagger .
\ee
An immediate consequence of this observation is that the energy spectrum is necessarily non-negative. Using the Pauli matrices, $\sigma_+ = \frac{1}{2}(\sigma_1+i\sigma_2)=\pmatrix{0&1\cr
0&0}~~;~~
\sigma_- =\frac{1}{2}(\sigma_1-i\sigma_2)= \pmatrix{0&0\cr
1&0}$, the operator ${\cal Q}$ can be further written as the sum ${\cal Q} = Q + {Q}^\dagger$, where 
\be
Q = \frac{i}{\sqrt{m}} \pi_- \sigma_- ~;~
Q^\dagger = \frac{-i}{\sqrt{m}} \pi_+ \sigma_+ 
\ee
are two complex, nilpotent, $Q^2=0=(Q^\dagger)^2$, supersymmetry charges. 
These charges, together with the Hamiltonian, obey the supersymmetry algebra\cite{Witten}-\cite{Khare}
\bea
\left\{ Q, Q\right\} = &0&=\left\{ {Q}^\dagger, {Q}^\dagger\right\}\cr
\left\{ Q, {Q}^\dagger\right\} &=& 2H \cr
\left[ Q, H \right] = &0& = \left[{Q}^\dagger, H\right] .
\eea

Next let us consider the $H_{\uparrow}$ and $H_{\downarrow}$ eigenvalue problems. The general normalizable $H_{\uparrow}$ eigenstate, $\psi_{\uparrow n}$, has energy $E_{\uparrow n} \ge 0$ and satisfies 
\be
H_{\uparrow}\psi_{\uparrow n}=\frac{\pi_+ \pi_-}{2m} \psi_{\uparrow n}=E_{\uparrow n}\psi_{\uparrow n}~~;~~E_{\uparrow n}\ge 0 .
\ee
Suppose $H_{\uparrow}$ has the normalizable zero energy ($E_{\uparrow 0}=0$) eigenstate
$\psi_{\uparrow 0}$ satisfying $H_{\uparrow}\psi_{\uparrow 0}=0$ which implies that $\pi_-\psi_{\uparrow 0}=0 $. The vanishing commutator of $Q$ and $Q^\dagger$ with $H$ implies that $\pi_- H_\uparrow = H_\downarrow \pi_- $ and $\pi_+ H_{\downarrow}=H_\uparrow \pi_+$. It follows that left multiplication of the $H_{\uparrow}$ eigenvalue equation by $\pi_-$ then dictates that $H_{\downarrow }(\pi_- \psi_{\uparrow n})=E_{\uparrow n} (\pi_- \psi_{\uparrow n})$, so that $\pi_- \psi_{\uparrow n}~~,~~ n>0,$ is an $H_{\downarrow}$ eigenstate with eigenvalue $E_{\uparrow n} > 0$. Note that the case $n=0$ does not give an $H_{\downarrow}$ eigenstate since $\pi_- \psi_{\uparrow 0}=0$. Thus the normalizable ground state of $H_{\downarrow}$ is $\psi_{\downarrow 0}=N_{\downarrow 0}\pi_- \psi_{\uparrow 1}$, where $N_{\downarrow 0}$ is a normalization constant, and has energy $E_{\uparrow 1}$. Except for the zero energy eigenstate of $H_{\uparrow}$, all the other eigenstates of $H_{\uparrow}$ and $H_{\downarrow}$ pair up with the same positive energy eigenvalues. This is a direct consequence of the supersymmetry. Thus if $\psi_{\uparrow n}$ is an eigenstate of $H_{\uparrow}$ with eigenvalue $E_{\uparrow n}$, then $\psi_{\downarrow n} = N_{\downarrow n} \pi_- \psi_{\uparrow n+1}~~,~~ n \ge 0,$ is an eigenstate of $H_{\downarrow}$ with eigenvalue $E_{\downarrow n}=E_{\uparrow n+1}>0$. 

To explicitly construct the zero energy ground state of $H_\uparrow$, we use that 
\be 
H_{\uparrow}\psi_{\uparrow 0}=0 
\ee
if and only if 
\be
\pi_- \psi_{\uparrow 0}=(\frac{\hbar}{i}\partial_-+\frac{ig}{c}\partial_- K)\psi_{\uparrow 0}=0 ~.
\ee
The solution is readily secured as 
\be
\psi_{\uparrow 0}(x_+,x_-)=T_-(e^{\frac{g}{\hbar c}\int_{x_{0-}}^{x_-}dx_-^\prime \partial_-^\prime K(x_+,x_-^\prime)})U_\uparrow ^\alpha (x_+) \eta_\alpha
\ee
where $\alpha$ is summed from $1$ to the dimension of $L^a$. The $U_\uparrow ^\alpha (x_+)$ are arbitrary functions of $x_+$ and $\eta_\alpha$ is a ${\rm dim} ~L^a$ component group space spinor. The $x_-$ ordered exponential is defined as 
\bea
T_-(e^{\frac{g}{\hbar c}\int_{x_{0-}}^{x_-}dx_-^\prime \partial_-^\prime K(x_+,x_-^\prime)})
&=&\sum_{n=0}^\infty \frac{1}{n!}(\frac{g}{\hbar c})^n \int_{x_{0-}}^{x_-} dx_{1-}\int_{x_{0-}}^{x_-} dx_{2-}...\int_{x_{0-}}^{x_-} dx_{n-}\cr
&&T_-[\partial_{1-}K(x_+,x_{1-})\partial_{2-}K(x_+,x_{2-})...\partial_{n-}K(x_+,x_{n-})]\cr
&=&\sum_{n=0}^\infty (\frac{g}{\hbar c})^n \int_{x_{0-}}^{x_-} dx_{1-}\int_{x_{0-}}^{x_{1-}} dx_{2-}...\int_{x_{0-}}^{x_{(n-1)-}} dx_{n-}\cr
&&[\partial_{1-}K(x_+,x_{1-})\partial_{2-}K(x_+,x_{2-})...\partial_{n-}K(x_+,x_{n-})]
\eea
with $K(x_+,x_{0-})=0$. This $H_{\uparrow}$ zero energy eigenstate will be the system ground state provided it is normalizable. 

On the other hand, suppose the $H_{\downarrow}$ eigenvalue equation
\be
H_{\downarrow}\psi_{\downarrow n}=E_{\downarrow n}\psi_{\downarrow n}~~;~~
E_{\downarrow n} \ge 0
\ee
admits the normalizable zero energy eigenstate, $\psi_{\downarrow 0}$, satisfying $\pi_+ \psi_{\downarrow 0} =0$ and $E_{\downarrow 0}=0$. Then an analogous argument gives the normalizable $H_{\uparrow}$ ground state as $\psi_{\uparrow 0}=N_{\uparrow 0} \pi_+ \psi_{\downarrow 1}$ having positive energy $E_{\uparrow 0}=E_{\downarrow 1}$. Once again, except for the zero energy eigenstate of $H_{\downarrow}$, all the other normalizable eigenstates of $H_{\downarrow}$ and $H_{\uparrow}$ pair up with degenerate positive energy eigenvalues. Thus  $\psi_{\uparrow n}=N_{\uparrow n}\pi_+ \psi_{\downarrow n+1}$ is an $H_{\uparrow}$ eigenstate with eigenvalue $E_{\uparrow n}=E_{\downarrow n+1}>0$. In this case, the zero energy ground state is gleaned from the condition 
\be
\pi_+ \psi_{\downarrow 0}=2(\frac{\hbar}{i}\partial_+-\frac{ig}{c}\partial_+ K)\psi_{\downarrow 0}=0
\ee
whose solution is
\be
\psi_{\downarrow 0}(x_+,x_-)=T_+(e^{-\frac{g}{\hbar c}\int_{x_{0+}}^{x_+}dx_+^\prime \partial_+^\prime K(x_+^\prime,x_{0+})})U_\downarrow ^\alpha (x_-) \eta_\alpha ~.
\ee
This $H_{\downarrow}$ zero energy eigenstate will be the system ground state provided it is normalizable.

If $K$ has only a single nonvanishing component in group space so that $K=L^N K^N$, then
\bea
T_-(e^{\frac{g}{\hbar c}\int_{x_{0_-}}^{x_{-}}dx_{-}^\prime \partial_{-}^\prime K(x_+,x_-^\prime)})
&=&e^{\frac{g}{\hbar c}L^N\int_{x_{0_-}}^{x_{-}} dx_{-}^\prime \partial_{-}^\prime K^N(x_+,x_-^\prime)} =e^{\frac{g}{\hbar c}L^N K^N(x_+,x_-)} \cr
T_+(e^{-\frac{g}{\hbar c}\int_{x_{0_+}}^{x_{+}}dx_{+}^\prime \partial_{+}^\prime K(x_+^\prime,x_-)})
&=&e^{-\frac{g}{\hbar c}L^N\int_{x_{0_+}}^{x_{+}} dx_{+}^\prime \partial_{+}^\prime K^N(x_+^\prime,x_-)} =e^{-\frac{g}{\hbar c}L^N K^N(x_+,x_-)}
\eea
with $K(x_+,x_{0_{-}})=0=K(x_{0_{+}},x_-)$ and
\be
\psi_{\uparrow 0}(x_+,x_-)=e^{\frac{g}{\hbar c}L^N K_N(x_+,x_-)}U_\uparrow^\alpha (x_+)\eta_\alpha
\ee
\be
\psi_{\downarrow 0}(x_+,x_-)=e^{-\frac{g}{\hbar c}L^N K_N(x_+,x_-)}U_\downarrow^\alpha (x_-) \eta_\alpha ~.
\ee
As an example, consider the case of the gauge group $SU(2)$. The fundamental representation matrices are simply $L^a=\frac{\tau^a}{2}$ with $\tau^a$ being the Pauli matrices. Taking $\eta_1 =\eta_\uparrow = \left[\begin{array}{rr}1\\0\end{array}\right]$ and $\eta_2 =\eta_\downarrow = \left[\begin{array}{rr}0\\1\end{array}\right]$ as the eigenstates of $L^3=\frac{\tau^3}{2}$ then 
\bea
\psi_{\uparrow 0}(x_+,x_-)&=&e^{\frac{g}{\hbar c} K^3(x_+,x_-)\frac{\tau^3}{2}}U_\uparrow^\alpha (x_+)\eta_\alpha \cr
&=&e^{\frac{g}{\hbar c} K^3(x_+,x_-)}U_\uparrow^\uparrow(x_+)\left[\begin{array}{rr}1\\0\end{array}\right]+
e^{-\frac{g}{\hbar c}K^3(x_+,x_-)}U_\uparrow^\downarrow(x_+)\left[\begin{array}{rr}0\\1\end{array}\right] ~.
\eea

For a uniform $B$ field, we can choose $K^3=-\frac{1}{2}B_0 y^2$ (asymmetric gauge) with $B_0 > 0$. In this case, only the isospin up gives a normalizable wavefunction so we set $U_\uparrow^\downarrow(x_+)=0$ yielding
\be
\psi_{\uparrow 0}(x_+,x_-)=e^{-\frac{g}{4\hbar c} B_0 y^2} U_\uparrow^\uparrow(x_+)\left[\begin{array}{rr}1\\0\end{array}\right] ~.
\ee
Note that the ground state has spin up and isospin up.  Since any function can be expanded in terms of plane waves, we can choose 
$U_\uparrow^\uparrow(x_+)=e^{ikx_+}$ and write
\be
\psi_{\uparrow 0 k}(x,y) =N_{\uparrow 0 k} e^{ik(x+iy)}e^{-\frac{gB}{4\hbar c}y^2}\left[\begin{array}{rr}1\\0\end{array}\right]
\ee
where we have included a label $k$ on the wavefunction which labels the degeneracy.

Following an analogous procedure, the zero energy eigenstate of $H_\downarrow$ is found as 
\bea
\psi_{\downarrow 0}(x_+,x_-)&=&e^{-\frac{g}{\hbar c} K^3(x_+,x_-)\frac{\tau^3}{2}}U_\downarrow^\alpha (x_-)\eta_\alpha \cr
&=&e^{-\frac{g}{\hbar c} K^3(x_+,x_)}U_\downarrow^\uparrow(x_-)\left[\begin{array}{rr}1\\0\end{array}\right]+
e^{\frac{g}{\hbar c}K^3(x_+,x_)}U_\downarrow^\downarrow(x_-)\left[\begin{array}{rr}0\\1\end{array}\right] ~.
\eea
For the uniform $B$ field case, only the isospin down yields a normalizable wavefunction so we set $U_\downarrow^\uparrow(x_-)=0$ and thus secure
\be
\psi_{\downarrow~0}(x_+,x_-)=e^{-\frac{g}{4\hbar c} B_0 y^2} U_\downarrow^\downarrow(x_-)\left[\begin{array}{rr}0\\1\end{array}\right] ~.
\ee
This time it is the spin down, isospin down state which is normalizable. Choosing 
$U_\downarrow^\downarrow(x_-)=e^{ikx_-}$ gives
\be
\psi_{\downarrow 0 k}(x,y) =N_{\downarrow 0 k} e^{ik(x-iy)}e^{-\frac{gB}{4\hbar c}y^2}\left[\begin{array}{rr}0\\1\end{array}\right] ~.
\ee
Note that the spin up, isospin up and spin down, isospin down states are degenerate zero energy ground states. 

Another example for the gauge group $SU(2)$ is provided by a vortex solution\cite{V}. Such a configuration arises from a prepotential with asymptotic form 
\be
K^1\sim 0~~;~~ K^2\sim 0~~;~~  K^3\sim -\frac{g_M}{4}\ell n \rho
\ee
with $g_M>0$. Here we have introduced the plane polar coordinates $\rho$ and $\varphi$ with $x_\pm =\rho e^{\pm i\varphi}$. The corresponding asymptotic vector potential components are
\bea
A_\rho &= &A_x cos \varphi +A_y sin \varphi \sim 0     \cr
A_\varphi &=& -A_x sin \varphi + A_y cos \varphi \sim \frac{\tau^3}{2}\frac{g_M}{4\rho}~.
\eea
while the resulting non-Abelian magnetic field strength vanishes asymptotically
\be
B(\rho, \varphi)\sim 0 .
\ee
Taking $U_{\uparrow}^{\uparrow}(x_+)=(x + iy)^n =\rho^n e^{in\varphi}~~,~~U_{\downarrow}^{\downarrow}(x_-)=(x -iy)^n =\rho^n e^{-n\varphi}$, and $U_{\downarrow}^{\uparrow}(x_-)=0~~;~~U_{\uparrow}^{\downarrow}(x_+)=0$ where $n$ a non-negative integer labeling the degeneracy, the normalizable zero energy ground states have the asymptotic form 
\be
\psi_{\uparrow 0}(\rho, \varphi)_n \sim N_{\uparrow 0~n}e^{in\varphi}
\rho^{n-\frac{g g_M}{8\hbar c}}\left[\begin{array}{rr}1\\0\end{array}\right], 
\ee
and 
\be
\psi_{\downarrow 0}(\rho, \varphi)_n \sim N_{\downarrow 0~n}e^{-in\varphi}
\rho^{n-\frac{g g_M}{8\hbar c}}\left[\begin{array}{rr}0\\1\end{array}\right] . 
\ee
Note that normalizability restricts $n < \frac{g g_M}{8\hbar c}-1$. Once again, there is a degeneracy between the spin up, isospin up and spin down, isospin down states.

As a final example, consider the magnetic field 
\be
B(x,y)= \partial_+ f(x_+) \partial_- f^*(x_-) (f(x_+)\tau_+ +f^*(x_-)\tau_- )-(\frac{16\hbar c}{g})^{1/3}\partial_+f(x_+)\partial_- f^*(x_-)\tau_3 ~,
\ee
where $f(x_+)$ and $f^*(x_-)$ are arbitrary functions. Such a non-Abelian magnetic field can be obtained from the prepotential 
\be
K(x_+, x_-) =(\frac{\hbar^2 c^2}{2g^2})^{1/3} \left(f(x_+)\tau_+ +f^*(x_-)\tau_-\right) +(\frac{\hbar c}{32g})^{1/3}f(x_+)f^*(x_-)\tau_3 ~.
\ee
The corresponding normalized zero energy spin up ground state takes the form 
\be
\psi_{\uparrow 0}(x_+, x_-) = N_{\uparrow 0} e^{-(\frac{\hbar c}{32g})^{1/3}f(x_+)f^*(x_-)}U^\downarrow _\uparrow (x_+) \left[\begin{array}{rr}0\\1\end{array}\right] ~. 
\ee
This time the normalized ground state is spin up but isospin down. Similarly, the degenerate normalized zero energy spin down ground state is 
\be
\psi_{\downarrow 0}(x_+, x_-) = N_{\downarrow 0} e^{-(\frac{\hbar c}{32g})^{1/3}f(x_+)f^*(x_-)}U^\uparrow _\downarrow (x_-) \left[\begin{array}{rr}1\\0\end{array}\right] ~ 
\ee
which has isospin up.
\\
\\

\noindent
This work was supported in part by the U.S. Department 
of Energy under grant DE-FG02-91ER40681 (Task B).

\end{document}